\documentclass[11pt]{article}
\usepackage{epsfig}

\def\la{\mathrel{\mathpalette\fun <}}
\def\ga{\mathrel{\mathpalette\fun >}}
\def\fun#1#2{\lower3.6pt\vbox{\baselineskip0pt\lineskip.9pt
  \ialign{$\mathsurround=0pt#1\hfil##\hfil$\crcr#2\crcr\sim\crcr}}}
\newcount\eLiNe\eLiNe=\inputlineno\advance\eLiNe by -1
\title{Some Current Theoretical Issues around Ultra-High Energy Cosmic Rays
\thanks{Send any remarks to {\tt sigl@iap.fr}}%
}

\author{G\"unter Sigl\\
GReCO, Institut d'Astrophysique de Paris, C.N.R.S.,\\
98 bis boulevard Arago, F-75014 Paris, France\\
F\'{e}d\'{e}ration de Recherche Astroparticule et Cosmologie,\\
Universit\'{e} Paris 7, 2 place Jussieu, F-75251 Paris Cedex 05, France}

\begin{document}

\maketitle

\begin{abstract}
We address some current theoretical issues around ultra-high energy
cosmic rays. We recall that scenarios producing more $\gamma-$rays
than cosmic rays up to high redshift can in general only provide
a sub-dominant contribution to the ultra-high energy cosmic ray
flux. This includes extra-galactic top-down and the Z-burst scenarios.
Finally we discuss the influence of large scale cosmic magnetic
fields on ultra-high energy cosmic ray propagation which is currently
hard to quantify. The views presented here represent the authors
perspective.
\end{abstract}

\section{Introduction}
High energy cosmic ray (CR) particles are shielded
by Earth's atmosphere and reveal their existence on the
ground only by indirect effects such as ionization and
showers of secondary charged particles covering areas up
to many km$^2$ for the highest energy particles. In fact,
in 1912 Victor Hess discovered CRs by measuring ionization from
a balloon~\cite{hess}, and in 1938 Pierre Auger proved the existence of
extensive air showers (EAS) caused by primary particles
with energies above $10^{15}\,$eV by simultaneously observing
the arrival of secondary particles in Geiger counters many meters
apart~\cite{auger_disc}.

After almost 90 years of research, the origin of cosmic rays
is still an open question, with a degree of uncertainty
increasing with energy~\cite{crbook}: Only below 100 MeV
kinetic energy, where the solar wind shields protons coming
from outside the solar system, the sun must give rise to
the observed proton flux. Above that energy the CR spectrum
exhibits little structure and is approximated
by broken power laws $\propto E^{-\gamma}$:
At the energy $E\simeq4\times 10^{15}\,$eV
called the ``knee'', the flux of particles per area, time, solid angle,
and energy steepens from a power law index $\gamma\simeq2.7$
to one of index $\simeq3.0$. The bulk of the CRs up to at least
that energy is believed to originate within the Milky Way Galaxy,
typically by shock acceleration in supernova remnants.
The spectrum continues with a further steepening to $\gamma\simeq3.3$
at $E\simeq4\times 10^{17}\,$eV, sometimes called the ``second knee''.
There are experimental indications that the chemical composition
changes from light, mostly protons, at the knee to domination by
iron and even heavier nuclei at the second knee~\cite{kascade}.
This is in fact expected in any scenario where acceleration and
propagation is due to magnetic fields whose effects only depend
on rigidity, the ratio of charge to rest mass, $Z/A$. This is true
as long as energy losses and interaction effects, which in general depend
on $Z$ and $A$ separately, are small, as is the case in the Galaxy, in
contrast to extra-galactic cosmic ray propagation at ultra-high energy.
Above the so called ``ankle'' or ``dip'' at $E\simeq5\times10^{18}\,$eV, the
spectrum flattens again to a power law of index $\gamma\simeq2.8$.
This latter feature
is often interpreted as a cross over from a steeper Galactic
component, which above the ankle cannot be confined by the Galactic
magnetic field, to a harder component of extragalactic origin.
The dip at $E\simeq5\times10^{18}\,$eV could also be partially due
to pair production by extra-galactic protons, especially
if the extra-galactic component already starts to dominate below
the ankle, for example, around the second-knee~\cite{bgh}. This
latter possibility appears, however, less likely in light of
a rather heavy composition up to the ankle suggested by several
experiments~\cite{kascade}. In any case, an eventual cross over
to an extra-galactic component is also in line with experimental
indications for a chemical composition becoming again lighter above
the ankle, although a significant heavy component is not
excluded and the inferred chemical composition above
$\sim10^{18}\,$eV is sensitive to the model of air shower interactions
and consequently uncertain presently~\cite{watson}.
In the following we will restrict our discussion on ultra-high
energy cosmic rays (UHECRs) above the ankle.

Although statistically meaningful information about the UHECR energy
spectrum and arrival direction distribution has been accumulated, no
conclusive picture for the nature and distribution of the sources
emerges naturally from the data. There is on the one hand the approximate
isotropic arrival direction distribution~\cite{bm} which indicates that we are
observing a large number of weak or distant sources. On the other hand,
there are also indications which point more towards a small number of
local and therefore bright sources, especially at the highest energies:
First, the AGASA ground array claims statistically significant multi-plets of
events from the same directions within a few degrees~\cite{teshima1,bm},
although this is controversial~\cite{fw} and has not been seen so far
by the fluorescence experiment HiRes~\cite{finley}.
The spectrum of this clustered component is $\propto E^{-1.8}$ and thus
much harder than the total spectrum~\cite{teshima1}.
Second, nucleons above $\simeq70\,$EeV suffer heavy energy losses due to
photo-pion production on the cosmic microwave background
--- the Greisen-Zatsepin-Kuzmin (GZK) effect~\cite{gzk} ---
which limits the distance to possible sources to less than
$\simeq100\,$Mpc~\cite{stecker}. For a uniform source distribution
this would predict a ``GZK cutoff'', a drop in the spectrum.
However, the existence of this ``cutoff'' is not established yet
from the observations~\cite{bergman} and may even depend on the
part of the sky one is looking at: The ``cutoff' could be mitigated
in the northern hemisphere where more nearby accelerators related
to the local supercluster can be expected. Apart from the SUGAR array
which was active from 1968 until 1979 in Australia, all UHECR detectors
completed up to the present were situated in the northern hemisphere.
Nevertheless the situation is unclear even there: Whereas a cut-off
seems consistent with the few events above $10^{20}\,$eV recorded
by the fluorescence detector HiRes~\cite{hires}, it is not compatible
with the 8 events above $10^{20}\,$eV measured by the AGASA ground
array~\cite{agasa}. It can be remarked, however, that analysis of
data based on a single fluorescence telescope, the so-called
monocular mode in which most of the HiRes data were obtained, is complicated
due to atmospheric conditions varying from event to event~\cite{cronin}.
The solution of this problem may have to await more analysis and,
in particular, the completion of the Pierre Auger project~\cite{auger}
which will combine the two complementary detection techniques
adopted by the aforementioned experiments and whose southern site
is currently in construction in Argentina.

This currently unclear experimental situation could easily be solved if it
would be possible to follow the UHECR trajectories backwards to their
sources. However, this may be complicated by the possible presence of
extragalactic magnetic fields, which would deflect the particles during
their travel. Furthermore, since the GZK-energy losses are of stochastic
nature, even a detailed knowledge of the extragalactic magnetic fields would
not necessarily allow to follow a UHECR trajectory backwards to its source
since the energy and therefor the Larmor radius of the particles
have changed in an
unknown way. Therefore it is not clear if charged particle astronomy with
UHECRs is possible in principle or not. And even if possible, it remains
unclear to which degree the angular 
resolution would be limited by magnetic deflection. This topic
will be discussed in Sect.~3.

The physics and astrophysics of UHECRs are also intimately linked with
the emerging field of neutrino astronomy (for reviews see
Refs.~\cite{nu_review}) as well as with the already
established field of $\gamma-$ray astronomy (for reviews see, e.g.,
Ref.~\cite{gammarev}). Indeed, all
scenarios of UHECR origin, including the top-down models, are severely
constrained by neutrino and $\gamma-$ray observations and limits.
In turn, this linkage has important consequences for theoretical
predictions of fluxes of extragalactic neutrinos above about a TeV
whose detection is a major goal of next-generation
neutrino telescopes: If these neutrinos are
produced as secondaries of protons accelerated in astrophysical
sources and if these protons are not absorbed in the sources,
but rather contribute to the UHECR flux observed, then
the energy content in the neutrino flux can not be higher
than the one in UHECRs, leading to the so called Waxman-Bahcall
bound for transparent sources with soft acceleration
spectra~\cite{wb-bound,mpr}.
If one of these assumptions does not apply, such as for acceleration
sources with injection spectra harder than $E^{-2}$ and/or opaque
to nucleons, or in the top-down scenarios where X particle decays
produce much fewer nucleons than $\gamma-$rays and neutrinos,
the Waxman-Bahcall bound does not apply, but the neutrino
flux is still constrained by the observed diffuse $\gamma-$ray
flux in the GeV range. This will be discussed in the following
section.

\section{Severe Constraints on Scenarios producing more photons
than hadrons}

Electromagnetic (EM) energy injected above the threshold for
pair production on the cosmic microwave background (CMB) at
$\sim10^{15}/(1+z)\,$eV at redshift $z$ (to a lesser extent
also on the infrared/optical background, with lower threshold) leads
to an EM cascade, an interplay between pair production followed
by inverse Compton scattering of the produced electrons. This
cascade continues until the photons fall below the pair production
threshold at which point the universe becomes transparent for them.
In todays universe this happens within just a few Mpc for injection
up to the highest energies above $10^{20}\,$eV. All EM energy
injected above $\sim10^{15}\,$eV and at distances beyond a few
Mpc today is therefore recycled to lower energies where it gives
rise to a characteristic cascade spectrum $\propto E^{-2.1}$ down
to fractions of a GeV~\cite{bere}. The universe thus acts as a calorimeter
where the total EM energy injected above $\sim10^{15}/(1+z)\,$eV
is measured as a diffuse isotropic $\gamma-$ray flux in the GeV regime.
This diffuse flux is not very sensitive to the somewhat uncertain
infrared/optical background~\cite{ahacoppi}.
Any observed diffuse $\gamma-$ray background acts as an upper limit
on the total EM injection.
Since in any scenario involving pion production the EM energy fluence
is comparable to the neutrino energy fluence, the constraint
on EM energy injection also constrains allowed neutrino fluxes.

This diffuse extragalactic GeV $\gamma-$ray background can be
extracted from the total $\gamma-$ray flux measured by EGRET by
subtracting the Galactic contribution. Since publication of the
original EGRET limit in 1995~\cite{egret}, models for this high
latitude Galactic $\gamma-$ray foreground were improved
significantly. This allowed the authors of Ref.~\cite{egret_new}
to reanalyze limits on the diffuse extragalactic background
in the region 30 MeV-10 GeV and to lower it by a factor 1.5-1.8
in the region around 1 GeV. There are even lower
estimates of the extragalactic diffuse $\gamma-$ray flux~\cite{kwl}.
In this article, however, we will use the more conservative 
limits from Ref.\cite{egret_new}.

The energy in the extra-galactic $\gamma-$ray background estimated
in Ref.~\cite{egret_new} is slightly more than one hundred times the
energy in UHECR above the GZK cutoff. The range of such trans-GZK
cosmic rays is about $\simeq30\,$Mpc, roughly one hundredth the
Hubble radius, and only sources within that GZK range contribute
to the trans-GZK cosmic rays. Therefore, any mechanism involving
sources distributed roughly uniformly on scales of the GZK energy
loss length $\simeq30\,$Mpc and producing a comparable amount of energy
in trans-GZK cosmic rays and photons above the pair production threshold
can potentially explain this energy flux ratio. The details depend
on the exact redshift dependence of source activity and other
parameters and in general have to be verified by numerically solving
the relevant transport equations, see, e.g., Ref.~\cite{ss}. Such
mechanisms include shock acceleration in powerful objects such as
active galactic nuclei~\cite{ta}.

On the other hand, any mechanism producing considerably {\it more}
energy in the EM channel above the pair production threshold than
in trans-GZK cosmic rays tend to predict a ratio of the diffuse
GeV $\gamma-$ray flux to the trans-GZK cosmic ray flux too high
to explain both fluxes at the same time. As a consequence, if
normalized at or below the observational GeV $\gamma-$ray background, such
scenarios tend to explain at most a fraction of the observed
trans-GZK cosmic ray flux. Such scenarios include particle physics
mechanisms involving pion production by quark fragmentation, e.g.
extra-galactic top-down mechanisms where UHECRs are produced by
fragmenting quarks resulting from decay of superheavy relics~\cite{bs-rev}.
Most of these quarks would fragment into pions rather than nucleons
such that more $\gamma-$rays (and neutrinos) than cosmic rays
are produced. Overproduction of GeV $\gamma-$rays can be avoided
by assuming the sources in an extended Galactic halo with a high $\ga10^3$
overdensity compared to the average cosmological source density, which
would also avoid the GZK cutoff~\cite{bkv}.
These scenarios, however, start to be constrained by the anisotropy
they predict because of the asymmetric position of the Sun
in the Galactic halo for which there are no indications in present
data~\cite{ks2003}. Scenarios based on quark fragmentation also become
problematic in view of a possible heavy
nucleus component and of upper limits on the photon fraction of
the UHECR flux~\cite{watson}.

\begin{figure}[ht]
\begin{center}
\includegraphics[angle=270,width=.98\textwidth,clip=true]{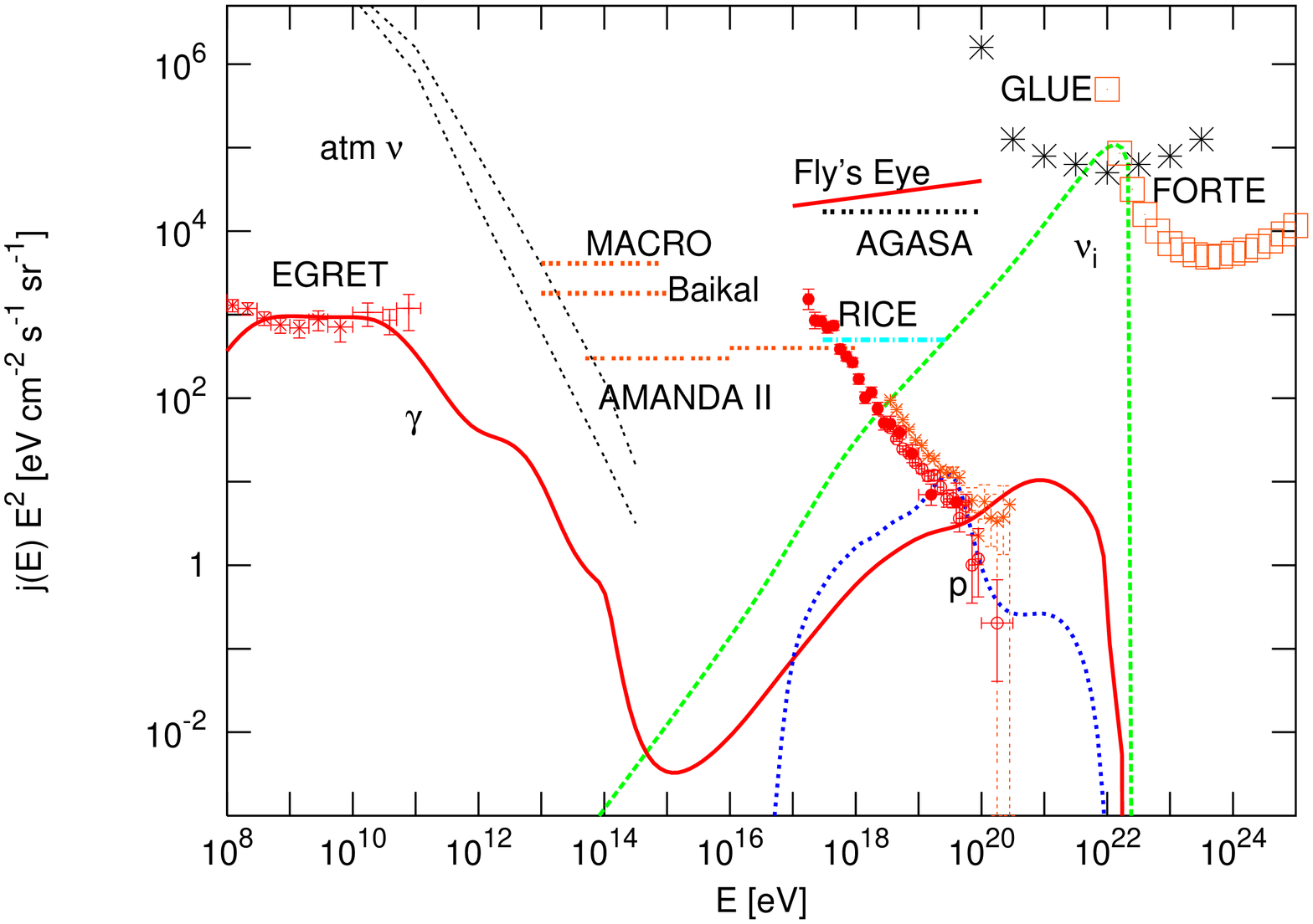}
\end{center}
\caption[...]{Flux predictions for a Z-burst model averaged over flavors and
characterized by a neutrino injection flux per comoving volume
$\propto E^{-1}$ up to $3\times10^{22}\,$eV and for redshifts between
0 and 3.
The sources are assumed to be exclusive neutrino emitters. All
neutrino masses were assumed equal with $m_\nu=0.33$~eV and we again
assumed maximal mixing between all flavors. Also shown are
predicted and observed cosmic ray and $\gamma-$ray fluxes, the
atmospheric neutrino flux~\cite{lipari}, as well as existing upper
limits on the diffuse neutrino fluxes from MACRO~\cite{MACRO},
AMANDA II~\cite{amandaII}, BAIKAL~\cite{baikal_limit},
AGASA~\cite{agasa_nu}, the Fly's Eye~\cite{baltrusaitis} and
RICE~\cite{rice_new} experiments, and the limits obtained with the
Goldstone radio telescope (GLUE)~\cite{glue} and the FORTE
satellite~\cite{forte}, as indicated. The cosmic ray data are
from the AGASA~\cite{agasa} and HiRes~\cite{hires} experiments, and
the new EGRET estimate of the extra-galactic diffuse $\gamma-$ray
flux is shown to the left. From Ref.~\cite{ss}.
\label{fig1}}
\end{figure}

As a specific example for scenarios involving quark fragmentation,
we consider here the case of decaying
Z-bosons. In this ``Z-burst mechanism'' Z-bosons are produced by
UHE neutrinos interacting with the relic neutrino
background~\cite{zburst1}. If the relic neutrinos
have a mass $m_\nu$, Z-bosons can be resonantly produced by UHE
neutrinos of energy
$E_\nu\simeq M_Z^2/(2m_\nu)\simeq4.2\times10^{21}\,{\rm eV}\,({\rm eV}/m_\nu)$.
The required neutrino
beams could be produced as secondaries of protons accelerated
in high-redshift sources. The fluxes predicted in these scenarios
have recently been discussed in detail, for example, in Refs.~\cite{fkr,ss}.
In Fig.~\ref{fig1} we show an optimistic example taken from Ref.~\cite{ss}.
It is assumed that the relic neutrino background has no significant
local overdensity. Furthermore, the sources
are assumed to not emit any $\gamma-$rays, otherwise the Z-burst
model with acceleration sources overproduces the diffuse GeV $\gamma-$ray
background~\cite{kkss}. We note that no known
astrophysical accelerator exists that meets the requirements
of the Z-burst model~\cite{kkss,gtt2003}.

However, a combination of new constraints allows to rule out
that the Z-burst mechanism explains a dominant fraction of the
observed UHECR flux, even for pure neutrino emitting sources:
A combination of cosmological data including the WMAP experiment limit
the sum of the masses of active neutrinos to $\la1\,$eV~\cite{hannestad}.
Solar and atmospheric neutrino oscillations indicate that individual
neutrino masses are nearly degenerate on this scale~\cite{mstv}, and thus
the neutrino mass per flavor must satisfy $m_\nu\la0.33\,$eV.
However, for such masses phase space constraints limit the possible
over-density of neutrinos in our Local Group of galaxies to
$\la10$ on a length scale of $\sim1\,$Mpc~\cite{sm}. Since
this is considerably smaller than the relevant UHECR loss lengths,
neutrino clustering will not significantly reduce the necessary UHE neutrino
flux compared to the case of no clustering.
For the maximal possible value of the neutrino
mass $m_\nu \simeq 0.33\,$eV, the neutrino flux required for the Z-burst
mechanism to explain the UHECR flux is only in marginal conflict with the
FORTE upper limit~\cite{forte}, and factor 2 higher than the new GLUE
limit~\cite{glue}, as shown in Fig.~\ref{fig1}. For all other
cases the conflict with both the GLUE and FORTE limits is
considerably more severe. 
Also note that this argument does not depend on the shape of the low energy
tail of the primary neutrino spectrum which could thus be even
mono-energetic, as could occur in exclusive tree level decays
of superheavy particles into neutrinos~\cite{gk}. However, in addition
this possibility has been ruled out by overproduction of GeV
$\gamma-$rays due to loop effects in these particle decays~\cite{bko}.

The possibility that the
observed UHECR flux is explained by the Z burst scenario involving
normal astrophysical sources which produce both neutrinos and photons
by pion production is already ruled out by the former EGRET limit:
In this case the GeV $\gamma-$ray flux level would have roughly
the height of the peak of the neutrino flux multiplied with the
squared energy in Fig.~\ref{fig1}, thus a factor $\sim100$ higher
than the EGRET level.

Any further reduction in the estimated contribution of the true diffuse
extra-galactic $\gamma-$ray background to the observed flux, therefore,
leads to more severe constraints on the total EM injection.
For example, future $\gamma-$ray detectors such as GLAST~\cite{glast}
will test whether the diffuse extragalactic GeV $\gamma-$ray background
is truly diffuse or partly consists of discrete sources that could
not be resolved by EGRET. Astrophysical discrete contributions such
as from intergalactic shocks are in fact expected~\cite{astrocontr}.
This could further improve the cascade limit
to the point where even acceleration scenarios may become seriously
constrained.

\section{Cosmic Magnetic Fields and Their Influence on Ultra-High
Energy Cosmic Ray Propagation}

Cosmic magnetic fields are inextricably linked with cosmic rays
in several respects. First, they play a central role in Fermi
shock acceleration. Second, large scale extra-galactic magnetic
fields (EGMF) can cause significant deflection of charged cosmic
rays during propagation and thus obviously complicate the relation
between observed UHECR distributions and their sources.

Magnetic fields are omnipresent in the Universe, but their
true origin is still unclear~\cite{bt_review}. Magnetic fields
in galaxies are observed with typical strengths of a few
micro Gauss, but there are also some indications for fields correlated
with larger structures such as galaxy clusters~\cite{bo_review}.
Magnetic fields as strong as
$\simeq 1 \mu G$ in sheets and filaments of the large scale galaxy
distribution, such as in our Local Supercluster, are compatible with
existing upper limits on Faraday rotation~\cite{bo_review,ryu,blasi}.
It is also possible that fossil cocoons of former radio galaxies,
so called radio ghosts, contribute significantly to the isotropization
of UHECR arrival directions~\cite{mte}.

To get an impression of typical deflection angles one can characterize the
EGMF by its r.m.s. strength $B$ and a coherence length $l_c$.
If we neglect energy loss processes for the moment, then
the r.m.s. deflection angle over a distance $r\ga l_c$ in such a field
is $\theta(E,r)\simeq(2rl_c/9)^{1/2}/r_L$~\cite{wm}, where the Larmor
radius of a particle of charge $Ze$ and energy $E$ is
$r_L\simeq E/(ZeB)$. In numbers this reads
\begin{equation}
  \theta(E,r)\simeq0.8^\circ\,
  Z\left(\frac{E}{10^{20}\,{\rm eV}}\right)^{-1}
  \left(\frac{r}{10\,{\rm Mpc}}\right)^{1/2}
  \left(\frac{l_c}{1\,{\rm Mpc}}\right)^{1/2}
  \left(\frac{B}{10^{-9}\,{\rm G}}\right)\,,\label{deflec}
\end{equation}
for $r\ga l_c$. This expression makes it immediately obvious
that fields of fractions of micro Gauss lead to strong deflection
even at the highest energies.
This goes along with a time delay $\tau(E,r)\simeq r\theta(E,d)^2/4
\simeq1.5\times10^3\,Z^2(E/10^{20}\,{\rm eV})^{-2}
(r/10\,{\rm Mpc})^{2}(l_c/{\rm Mpc})(B/10^{-9}\,{\rm G})^2\,$yr
which can be millions of years. A source visible in UHECRs today
could therefore be optically invisible since many models involving,
for example, active galaxies as UHECR accelerators, predict
variability on shorter time scales.

Quite a few simulations of the effect of extragalactic magnetic fields
(EGMF) on UHECRs exist in the literature, but usually idealizing
assumptions concerning properties and distributions of sources
or EGMF or both are made: In Refs.~\cite{slb,ils,lsb,sse,is} sources
and EGMF follow a pancake profile mimicking the local supergalactic
plane. In other studies EGMF have been approximated
in a number of fashions: as negligible~\cite{sommers,bdm},
as stochastic with uniform statistical properties~\cite{bo,ynts,ab},
or as organized in spatial cells with a given coherence length and a strength
depending as a power law on the local density~\cite{tanco}.
Only recently attempts have been made to simulate UHECR propagation
in a realistically structured universe~\cite{sme,dolag}. For
now, these simulations are limited to nucleons.

In Ref.~\cite{sme} the magnetized extragalactic environment used
for UHECR propagation is produced by a simulation of the large scale
structure of the Universe. The simulation was carried out
within a computational box of $50\,h^{-1}\,$Mpc length on a side, 
with normalized Hubble constant 
$h\equiv H_0/(100$ km s$^{-1}$ Mpc$^{-1})$ = 0.67, and using
a comoving grid of 512$^3$ zones and 256$^3$ dark matter
particles. The EGMF was initialized to zero at simulation start 
and subsequently its seeds were
generated at cosmic shocks through the Biermann battery
mechanism~\cite{kcor97}. Since cosmic shocks form
primarily around collapsing structures including filaments, the above
approach avoids generating EGMF in cosmic voids.

In Ref.~\cite{dolag} constrained simulations of the local large
scale structure were performed and the magnetic smoothed particle
hydrodynamics technique was used to follow EGMF evolution. The
EGMF was seeded by a uniform seed field of maximal strength compatible
with observed rotation measures in galaxy clusters.

The questions considered in these two works were somewhat different,
however. In Ref.~\cite{dolag} deflections of UHECR above
$4\times10^{19}\,$eV were computed as a function of the direction
to their source which were assumed to be at cosmological distances.
This made sense, because (i) the constrained simulations gives a
viable model of our local cosmic neighborhood within about 100 Mpc,
at least on scales beyond a few Mpc and (ii) the deflections typically
were found to be smaller than a few degrees. Concrete source distributions
were not considered.

\begin{figure}[ht]
\includegraphics[width=0.9\textwidth,clip=true]{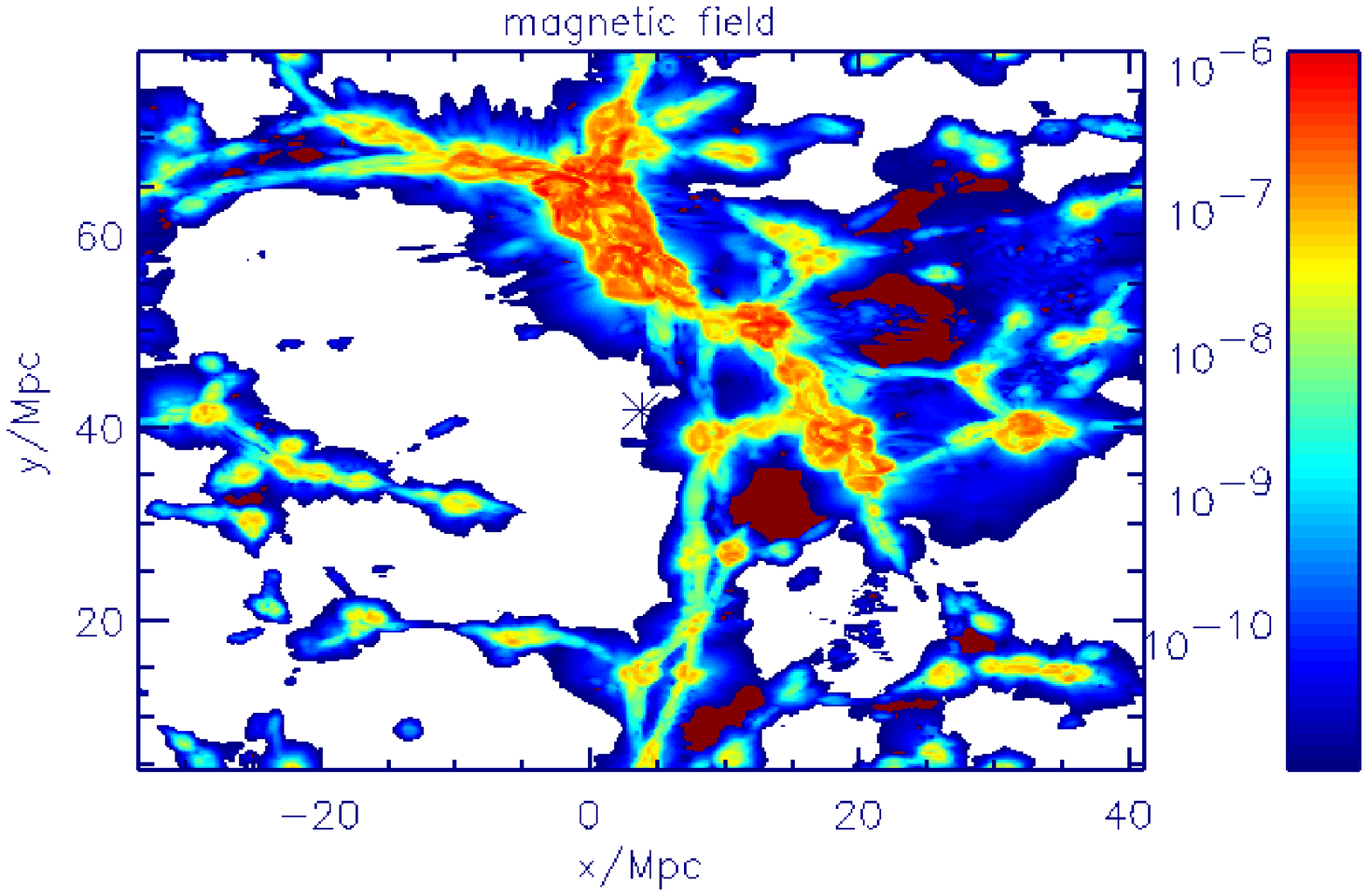}
\includegraphics[width=0.9\textwidth,clip=true]{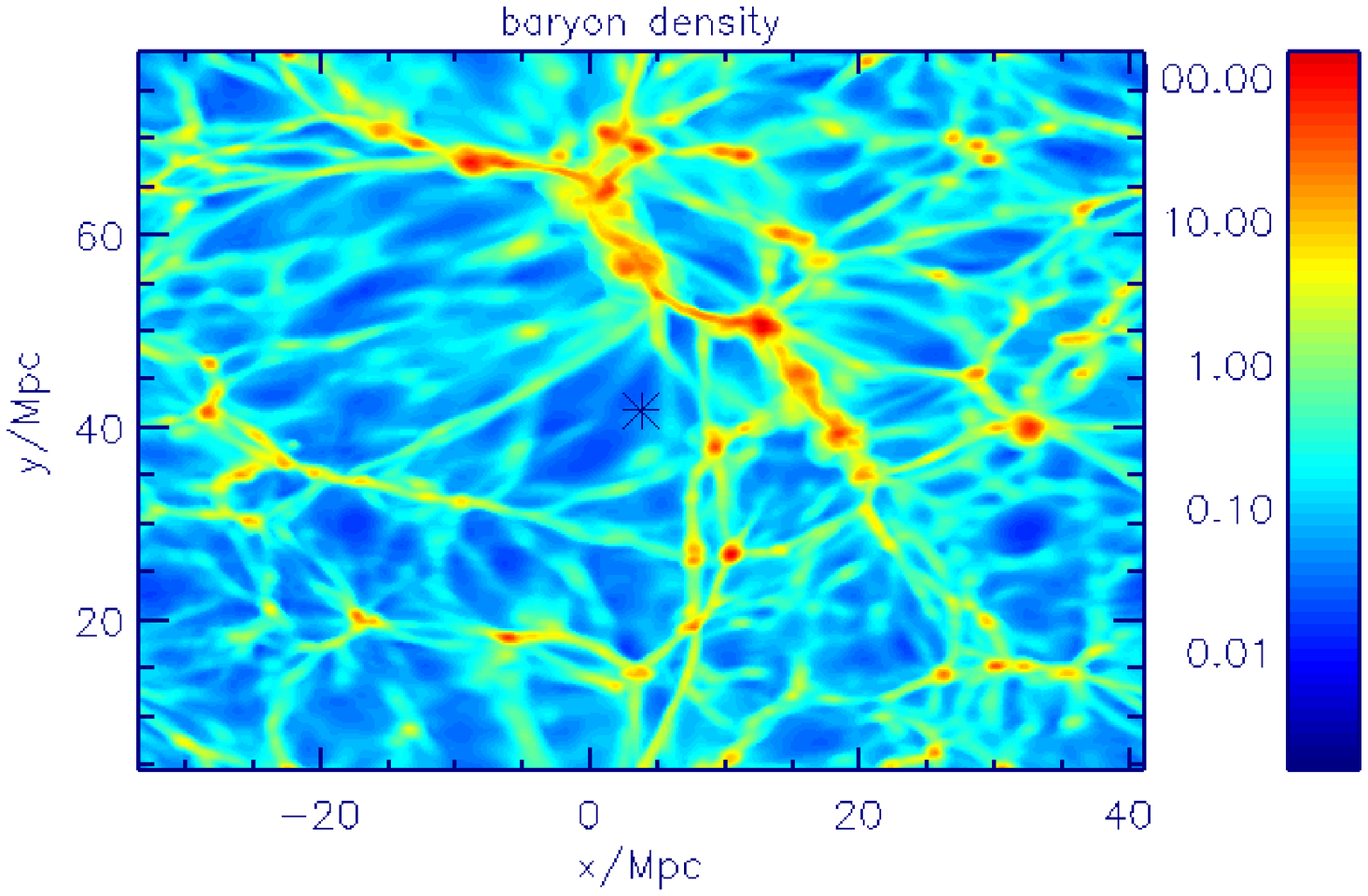}
\caption[...]{Log-scale two-dimensional cut through
magnetic field total strength in Gauss (color
scale in Gauss, upper panel) and baryon density in units of average
baryon density (color scale, lower panel), for a scenario
in good agreement with UHECR data studied in Ref.~\cite{sme}.
The observer is in the center of the figures and is marked by a star.
The EGMF strength at the observer is $\simeq10^{-11}\,$G. Note that
both panels correspond to the same cuts through the full large scale
simulation box used.}
\label{fig2}
\end{figure}

In contrast, Ref.~\cite{sme} was not concerned with concrete sky
distributions or deflection maps because the simulation was unconstrained
and thus only gave a typical large scale structure model and not our
concrete local neighborhood. Instead, the question was asked which
observer positions and source distributions and characteristics
lead to UHECR distributions whose spherical multi-poles for $l\leq10$
and auto-correlation at angles $\theta\la20^\circ$ are consistent
with observations. As a result it was found that (i) the observed
large scale UHECR isotropy requires the neighborhood within a few Mpc
of the observer is characterized by weak magnetic fields below $0.1\,\mu$G,
and (ii) once that choice is made, current data do not strongly
discriminate between uniform and structured source distributions
and between negligible and considerable deflection. Nevertheless,
current data moderately favor a scenario in which (iii) UHECR
sources have a density $n_s\sim10^{-5}\,{\rm Mpc}^{-3}$ and follow the matter
distribution and (iv) magnetic fields are relatively pervasive within the large
scale structure, including filaments, and with a strength of order of a $\mu$G
in galaxy clusters. A two-dimensional cut through the baryonic density
and EGMF environment of the observer in a typical such scenario is
shown in Fig.~\ref{fig2}.

It was also studied in Ref.~\cite{sme} how future data of considerably
increased statistics can be used to learn more about EGMF and source
characteristics. In particular, low auto-correlations at
degree scales imply magnetized sources quite independent of
other source characteristics such as their density. The latter can
only be estimated from the auto-correlations halfway reliably
if magnetic fields have negligible impact on propagation.
This is because if sources are immersed
in considerable magnetic fields, their images are smeared out,
which also smears out the auto-correlation function over several
degrees. For a sufficiently high source density, individual images
can thus overlap and sensitivity to source density is consequently
lost. The statistics expected from next generation experiments
such as Pierre Auger~\cite{auger} and EUSO~\cite{euso} should
be sufficient to test source magnetization by the auto-correlation
function~\cite{sme}.

\begin{figure}[ht]
\includegraphics[width=0.98\textwidth,clip=true]{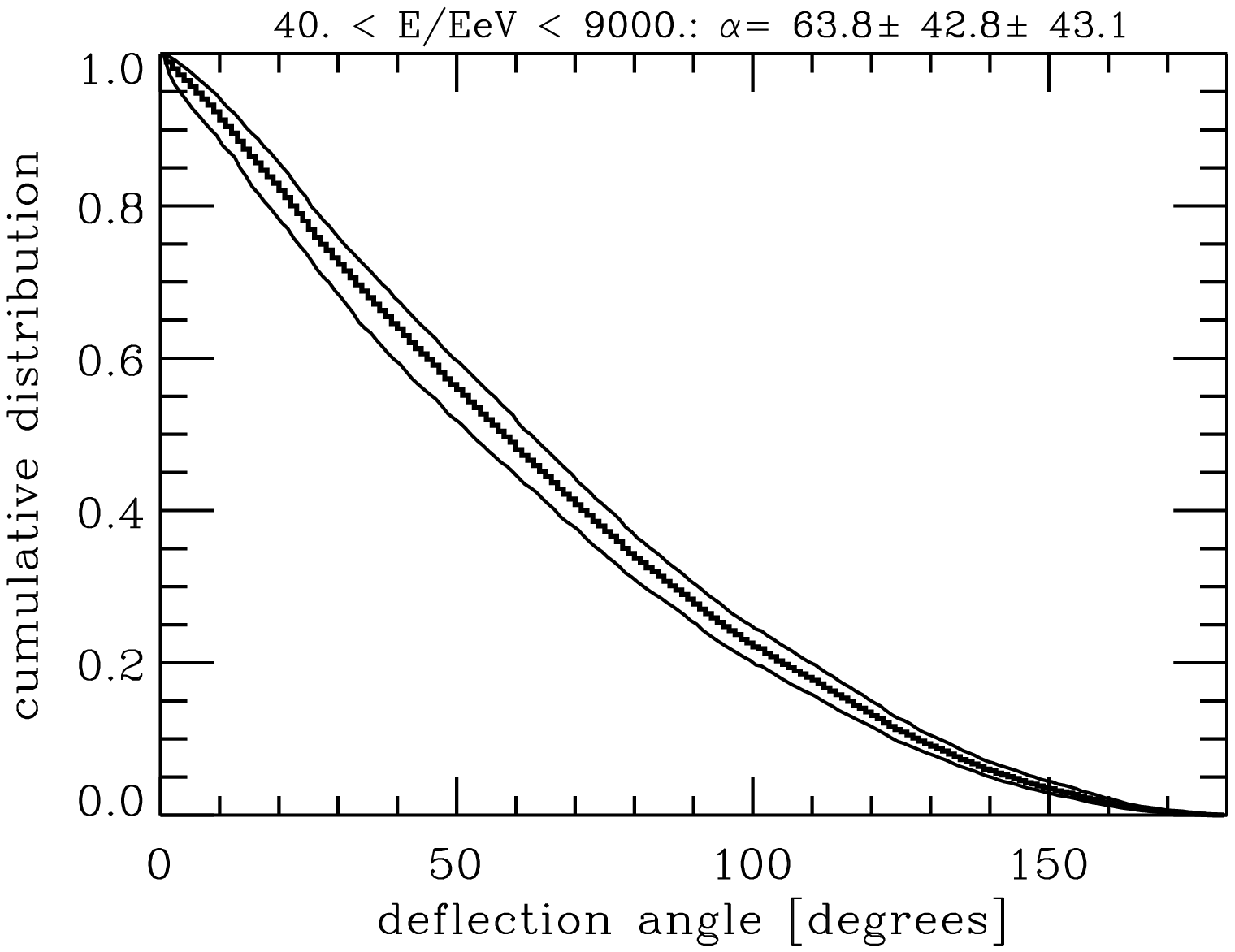}
\caption[...]{The cumulative distribution of minimal UHECR deflection
angles $\alpha$ with respect to the line of sight to the sources.
This is for a scenario from Ref.~\cite{sme} in good agreement with
UHECR data, where the sources follow the baryon density and have average
density $n_s=2.4\times10^{-5}\,{\rm Mpc}^{-3}$, and the EGMF included in
the large scale structure simulation reaches several micro Gauss in
the most prominent galaxy cluster. Shown are the average (middle,
histogram) and 1-$\sigma$ variations (upper and lower curves) above
$4\times10^{19}\,$eV, over 24 realizations varying in the positions
and luminosities $Q_i$ of individual sources,
the latter assumed to be distributed as $dn_s/dQ_i\propto Q_i^{-2.2}$
with $1\leq Q_i\leq100$ in arbitrary units. Also given on top
of the figure are average and variances of the distributions.}
\label{fig3}
\end{figure}

Interestingly, however, there is a considerable quantifiable difference
in the typical deflection angles predicted by the two EGMF scenarios
in Refs.~\cite{sme,dolag} that can {\it not} be compensated by
specific source distributions: Even for homogeneous source distributions,
the average deflection angle for UHECRs above $4\times10^{19}\,$eV
obtained in Ref.~\cite{sme} is much larger than in Ref.~\cite{dolag},
as can be seen in Fig.~\ref{fig3}. In fact, even if the magnetic field
strength is reduced by a factor 10 in the simulations of Ref.~\cite{sme},
the average deflection angle above $4\times10^{19}\,$eV is still
$\sim30^\circ$, only a factor $\simeq2.2$ smaller. This non-linear
behavior of deflection with field normalization is mostly due to the
strongly non-homogeneous character of the EGMF.

\begin{figure}[ht]
\includegraphics[width=0.98\textwidth,clip=true]{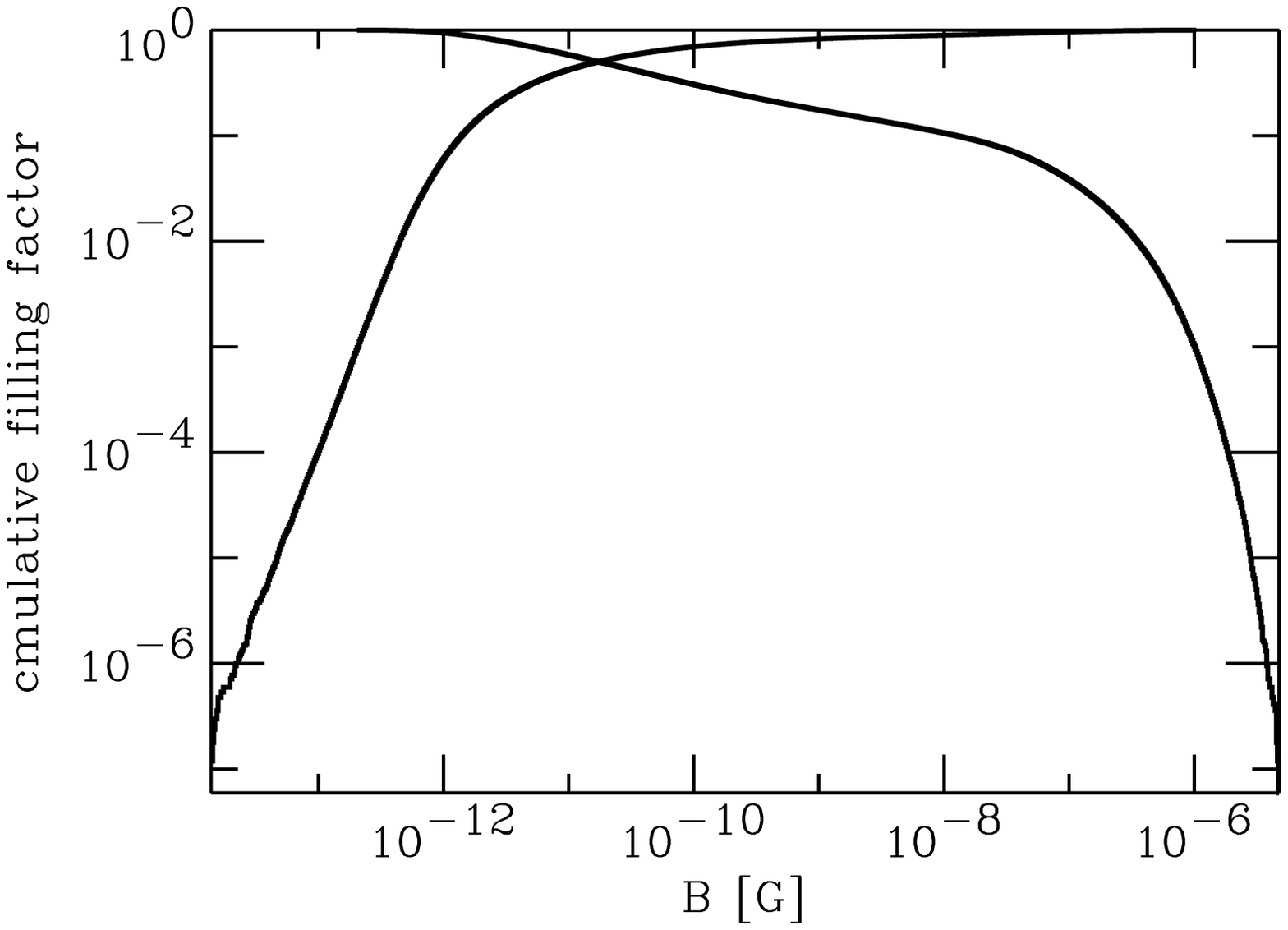}
\caption[...]{The cumulative filling factors for EGMF strength
in the simulations used in Ref.~\cite{sme} above (decreasing curve)
and below (increasing curve) a given threshold, as a function of
that threshold.}
\label{fig4}
\end{figure}

Most of these differences are probably due the different 
numerical models for the magnetic fields.
Although Ref.~\cite{dolag} start with uniform seed fields,
whereas in Ref.~\cite{sme} seed fields are injected
at shocks, by itself, this difference
should not influence the resulting EGMF very
much at late times, at least inside galaxy clusters~\cite{ryu}.
It should be noted, however, that in the filaments, where 
the gas motions are more uniform, the simulated magnetic fields may
depend to a certain extent on the initial seed fields although 
that is not trivial to quantify in general terms. 
In addition, numerical resolution
may play an important role because it affects the amplification and
the topological structure of the 
magnetic fields, both of which are important for the normalization
procedure, see below. The resolution
in Ref.~\cite{sme} is constant and much better in filaments and
voids but worse in the core of galaxy clusters than the (variable) 
resolution in Ref.~\cite{dolag}. If in both simulations the magnetic
fields are normalized to (or reproduce) the same ``observed'' values 
in the core of rich clusters then obviously their values 
in the filaments will be very different for the reasons outlined above.
This may partly explain why the contribution
of filaments to UHECR deflection is more important in Ref.~\cite{sme},
although a more detailed analysis and comparison are required to settle 
the issue. In any case, the magnetic fields obtained
in Ref.~\cite{sme} seem to be quite extended, as can be seen in
Fig.~\ref{fig4}: About 10\% of the volume is filled with fields
stronger than 10 nano Gauss, and a fraction of $10^{-3}$ is
filled by fields above a micro Gauss. The different amounts of
deflection obtained in the simulations of Refs.~\cite{sme,dolag} show
that the distribution of EGMF and their effects on UHECR propagation
are currently rather uncertain.

Finally we note that these studies should be extended to include
heavy nuclei~\cite{prepa} since there are indications that a fraction
as large as 80\% of iron nuclei may exist above $10^{19}\,$eV~\cite{watson}.
As a consequence, even in the EGMF scenario of Ref.~\cite{dolag}
deflections could be considerable and may not allow particle astronomy
along many lines of sight: The distribution of deflection angles in
Ref.~\cite{dolag} shows that deflections of protons above
$4\times10^{19}\,$eV of $\ga1^\circ$ cover a considerable fraction
of the sky. Suppression of deflection along typical lines of sight
by small filling factors of deflectors is thus unimportant in this
case. The deflection angle of any
nucleus at a given energy passing through such areas will therefore
be roughly proportional to its charge as long as energy loss
lengths are larger than a few tens of Mpc~\cite{bils}. Deflection angles of
$\sim20^\circ$ at $\sim4\times10^{19}\,$eV should thus be the rule
for iron nuclei. In
contrast to the contribution of our Galaxy to deflection which
can be of comparable size but may be corrected for within sufficiently
detailed models of the galactic field, the extra-galactic contribution
would be stochastic. Statistical methods are therefore likely to
be necessary to learn about UHECR source distributions and
characteristics. In addition, should a substantial heavy composition
be experimentally confirmed up to the highest energies, some sources would
have to be surprisingly nearby, within a few Mpc, otherwise only
low mass spallation products would survive propagation~\cite{er}.

The clustered component of the UHECR spectrum may play a key role
in this context and may be caused by discrete sources in directions
with small deflection. Since, apart from energy losses, cosmic rays
of same rigidity $Z/A$ are deflected similarly by cosmic magnetic
fields, one may expect that the composition of the clustered component
may become heavier with increasing energy. Indeed, in Ref.~\cite{teshima}
it was speculated that the AGASA clusters may be consistent with
consecutive He, Be-Mg, and Fe bumps.

\section{Conclusions}
We have reviewed two current issues in theoretical ultra-high energy
cosmic ray research.

The first one concerns constraints on scenarios
attempting to explain highest energy cosmic rays by extra-galactic
sources producing not only cosmic rays but also photons: Improved
data analysis and, in the future, improved data for example from
GLAST can considerably reduce estimates of the true extra-galactic
GeV $\gamma-$ray background which acts as a calorimeter of electromagnetic
energy injected above $\sim10^{15}/(1+z)\,$eV. Already current estimates
imply that scenarios producing considerably more photons than
hadrons, such as extra-galactic top-down scenarios and the Z-burst
mechanism, can not explain all of the highest energy cosmic ray flux.
A further reduced diffuse GeV $\gamma-$ray background will start to
constrain even normal acceleration scenarios.

As for the second issue we pointed out that the influence of large
scale cosmic magnetic fields on ultra-high energy cosmic ray propagation
is currently hard to quantify and may not allow to do ``particle
astronomy'' along most lines of sight, especially if a significant
heavy nucleus component is present above $10^{19}\,$eV. In this case
extensive Monte Carlo simulations including nuclei and based on
constrained large scale structure simulations will be necessary
to fully exploit data from future instruments such as the Pierre
Auger~\cite{auger} and EUSO projects~\cite{euso}.

\section*{Acknowledgements} The material presented here is based on
collaborations with Torsten En\ss lin, Francesco Miniati and
Dmitry Semikoz whom I would especially like to thank. Thanks to
Francesco also for suggestions on the manuscript. I also thank
Igor Tkachev for discussions on differences between our UHECR
simulations in a structured magnetized universe.

\end{document}